\documentclass[amsmath,amssymb,nofootinbib]{revtex4}

\usepackage{graphicx}
\usepackage{dcolumn}
\usepackage{bm}
\begin{document}

\title{Nuclear giant resonances\footnote{invited talk at
the 9th Asia - Pacific Physics Conference, 
October 25 -- 31, 2004 - Hanoi, Vietnam}}

\author{Nguyen Dinh Dang\footnote{Corresponding address: 
Cyclotron Center, RIKEN, 2-1 Hirosawa, Wako city, 
351-0198 Saitama, Japan. E-mail: dang@postman.riken.go.jp}}
\affiliation{%
1 - Cyclotron Center, The Institute of Physical and Chemical Research, 
2-1 Hirosawa, Wako, 351-0198 
Saitama, Japan \\
2 - Institute for Nuclear Science and Technique, Vietnam Atomic 
Energy Commission, Hanoi - Vietnam}%
\begin{abstract}
This talk presents the recent status  of theoretical and experimental 
studies  of giant resonances in nuclei with the emphasis  on: (1) 
charge-exchange Gamow-Teller resonance, (2) multiple-phonon 
resoanances, (3) giant dipole resonances in highly excited nuclei, and
(4) pygmy dipole resonances in neutron rich nuclei.
In particular, the description of these resonances 
within the framework of the phonon damping model 
is discussed in detail.
\end{abstract}

\maketitle

\hspace{-4mm} {\bf INTRODUCTION}
\vspace{2mm} 

Giant resonances (GR) are fundamental modes of nuclear excitations at high 
frequencies. The best-known one of them is the giant dipole resonance (GDR), 
which was observed in photo nuclear reactions 56 years ago and is 
described as the collective motion of protons against protons 
according to the simplest 
theoretical model by Goldhaber and Teller. The collective model of 
nucleus indicates that the nucleus should be studied in terms of normal 
modes, many of which are vibrational modes. Since the GDR is a giant 
vibration, by studying the GDR we learn a great deal about how the 
single-particle motion is coupled to vibrations, hence about the 
nuclear structure itself. Many other types of GR 
were measured later. They include giant multipole resonances such as 
the E0 giant monopole (GMR), E2 giant quadrupole (GQR), 
isoscalar E3 resonances seen in 
$(e,e')$ and $(\alpha,\alpha')$ reactions, 
magnetic M1 resonances GDR observed in $(p,p')$ reactions, charge 
-exchange Gamow-Teller resonance extracted in $(p,n)$ reactions. 
Recently the isoscalar GDR, the multiple-phonon GR, and the GDR in 
highly excited nuclei (hot GDR) were also observed. With the 
development of research in neutron-rich nuclei, new modes of 
excitations such as soft-dipole in neutron-halo nuclei, pygmy 
resonances in neutron-skin nuclei, and their coupling to GDR 
were also studied.

In this talk I will present a simple model, called phonon-damping 
model (PDM), which turns out to be successful in describing
simultaneously many of these resonances, including 
the GDR in hot nuclei, double GDR (DGDR), Gamow-Teller resonance 
(GTR) in stable nuclei, as well as pigmy dipole resonances 
(PDR) in neutron-rich nuclei.
\vspace{4mm} 

\hspace{-4mm} {\bf THE PHONON DAMPING MODEL}
\vspace{2mm} 

The PDM has been proposed in 1998 in Ref. 
\cite{PDM}, and developed further in a series of 
papers~\cite{series,PDR}. 
According to the PDM 
the propagation of the GR phonon is damped 
due to coupling to quasiparticle field. 
The final equation of the Green function for the GR propagation   
has the form~\cite{PDR}
\begin{equation}
G_{\lambda i}(E)=\frac{1}{2\pi}\frac{1}{E-\omega_{\lambda 
i}-P_{\lambda i}(E)}~,
\label{GE}
\end{equation}
where the explicit form of 
the polarization operator $P_{\lambda i}(E)$
is
\begin{equation}
P_{\lambda i}(E)=\frac{1}{\hat{\lambda}^{2}}\sum_{jj'}[f_{jj'}^{(\lambda)}]^{2}
\biggl[\frac{(u_{jj'}^{(+)})^{2}(1-n_{j}-n_{j'})(\epsilon_{j}+\epsilon_{j'})}
{E^{2}-(\epsilon_{j}+\epsilon_{j'})^{2}}
- \frac{(v_{jj'}^{(-)})^{2}(n_{j}-n_{j'})(\epsilon_{j}-\epsilon_{j'})}
{E^{2}-(\epsilon_{j}-\epsilon_{j'})^{2}}\biggr].
\label{PE}
\end{equation}
Here $u_{jj'}^{(+)}=u_{j}v_{j'}+u_{j'}v_{j}$, 
$v_{jj'}^{(-)}=u_{j}u_{j'}-v_{j}v_{j'}$ are combinations of Bogolyubov 
$(u,v)$ factors, $\epsilon_{j}$ are quasiparticle energies, and $n_{j}$ 
are the temperature-dependent quasiparticle-occupation numbers, whose 
form is close to that given by the Fermi-Dirac distribution.
The phonon damping $\gamma_{\lambda i}(\omega)$ ($\omega$ real) is 
obtained as the imaginary part of the analytic continuation of
the polarization operator $P_{\lambda i}(E)$ into the complex energy 
plane $E=\omega\pm i\varepsilon$:
\[
\gamma_{\lambda i}(\omega)=\frac{\pi}{2\hat{\lambda}^{2}}
\sum_{jj'}[f_{jj'}^{(\lambda)}]^{2}\biggl\{
(u_{jj'}^{(+)})^{2}(1-n_{j}-n_{j'})
[\delta(E-\epsilon_{j}-\epsilon_{j'})-
\delta(E+\epsilon_{j}+\epsilon_{j'})]-
\]
\begin{equation}
(v_{jj'}^{(-)})^{2}(n_{j}-n_{j'})[\delta(E-\epsilon_{j}+\epsilon_{j'})
-\delta(E+\epsilon_{j}-\epsilon_{j'})]\biggr\}.
\label{gamma}
\end{equation}
The energy $\bar{\omega}$ 
of giant resonance (damped collective phonon) is found as the pole of the 
Green's function (\ref{GE}):
\begin{equation}
\bar{\omega}-\omega_{\lambda i}-P_{\lambda i}(\bar{\omega})=0~.
\label{energy}
\end{equation}
The width $\Gamma_{\lambda}$ of giant resonance 
is calculated as twice of the damping 
$\gamma_{\lambda}(\omega)$ at $\omega=\bar{\omega}$, i.e.
\begin{equation}
\Gamma_{\lambda}=2\gamma_{\lambda}(\bar{\omega}),
\label{width}
\end{equation}
where $\lambda=$ 1 corresponds to the GDR. 
The line shape of the GDR is described by the strength function 
$S_{\rm GDR}(\omega)$, which is   
derived as:
\begin{equation}
S_{\rm GDR}(\omega)=\frac{1}{\pi}\frac{\gamma_{\rm GDR}(\omega)}
{(\omega-\bar{\omega})^{2}+\gamma_{\rm GDR}^{2}(\omega)}~.
\label{S}
\end{equation} 
\vspace{4mm}

\hspace{-4mm}
{\bf COMPARISON OF THEORETICAL PREDICTIONS WITH EXPERIMENTAL DATA}
\vspace{2mm}        

\begin{figure}                                                             
\includegraphics{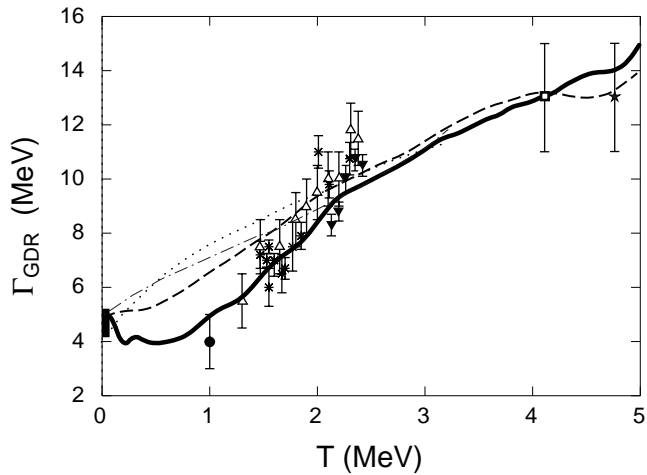}
\caption{\label{HOT}GDR width $\Gamma_{\rm GDR}$ as  a function of 
temperature $T$ for $^{120}$Sn. The dashed and solid lines show the 
PDM results obtained neglecting and including thermal pairing gap, 
respectively.
The predictions by two 
versions of the thermal shape-fluctuation model are shown as the 
dash-dotted~\cite{TFM} 
and thin dotted~\cite{Kuznesov} lines, respectively. The experimental 
data in taken from Refs. \cite{alpha}.}
\end{figure}

\begin{figure}                                                             
\includegraphics{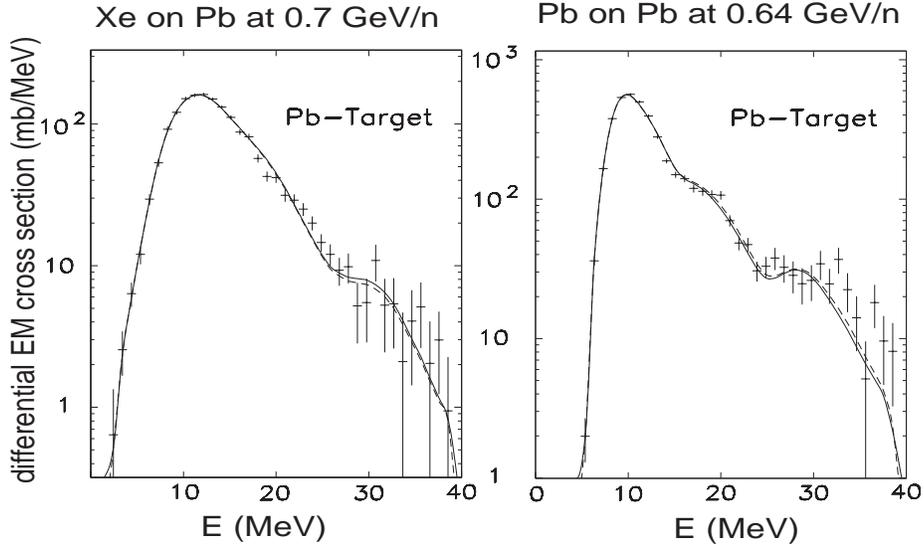}
\caption{EM cross sections of GDR and DGDR 
for $^{136}$Xe and $^{208}$Pb. The solid lines are theoretical 
predictions, in which the DGDR strength functions within PDM 
are used. The data points are results of 
the LAND collaboration~\cite{Bore}. The dashed lines show the best 
fit using $\chi^{2}$. The theoretical results have been folded with 
the detector response by K. Boretzky~\cite{Bore}.\label{DGDR}}
\end{figure}

The PDM has been proved to be quite successful in the 
description of the width and the shape of the GDR as a function of
temperature $T$ and angular momentum 
$J$. An example is shown in Fig. \ref{HOT}.
The PDM has resolved the long-standing problem with 
the electromagnetic (EM) cross sections of the DGDR in $^{136}$Xe
and $^{208}$Pb, in which the prediction by the non-interacting phonon 
picture underestimated significantly the observed DGDR cross sections 
by the LAND collaboration. The prediction using the strength functions
obtained within PDM~\cite{Au} is given in Fig. \ref{DGDR} in comparison with the 
latest results of data analyses by LAND collaboration~\cite{Bore}. The 
agreement between the PDM prediction and the data is remarkable. 

\begin{figure}                                                             
\includegraphics{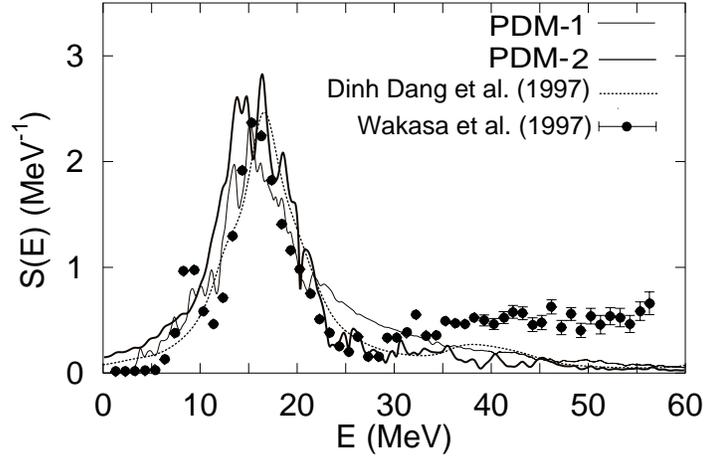}
\caption{Strength functions of the GTR in ${90}$Nb. See text for 
the notation.\label{GTR}}
\end{figure}

\begin{figure}                                                             
\includegraphics{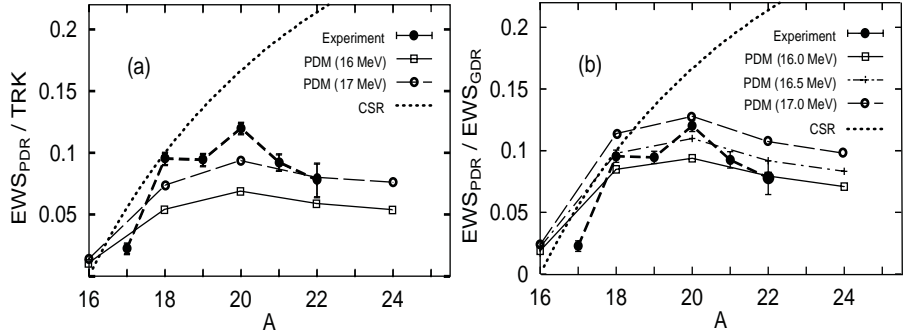}
\caption{EWS of PDR strength up to excitation energy $E_{\rm max}$ for
oxygen isotopes. Results obtained within PDM 
with $E_{\rm max}=$ 16, 16.5, and 
17 MeV are displayed as open boxes connected with solid line, 
crosses connected with dash-dotted line, and open circles connected 
with thin dashed line, respectively.  In (a) the PDM results 
are shown in units of Thomas-Reich-Kuhn sum rule (TRK), while in (b) they are in
units of the total GDR strength integrated up to 30 MeV. 
Experimental data (in units of TRK), obtained with $E_{\rm max}=$ 15 
MeV~\cite{GSI} are shown
by full circles connected with thick dashed line. The dotted line is
the prediction by the cluster sum rule (CSR) (in units of TRK). \label{EWSR_O}}
\end{figure}

Shown in Fig. \ref{GTR} is the prediction of \cite{DangGTR} within 
two versions of PDM, called PDM-1 
(thin solid line), and PDM-2 (thick solid line) (for the details see Refs. 
~\cite{PDM,series}) for the 
GTR in $^{90}$Nb in 
comparison with the result obtained within a microscopic theory which 
explicitly includes coupling to $2p2h$ configurations in terms of 
two-phonon configurations (dotted line)~\cite{DangGTR2p2h}, and 
the experimental data (data points with errorbars)~\cite{Wakasa}.
Again, the agreement between theory and experiment is quite reasonable.
\begin{figure}                                                             
\includegraphics{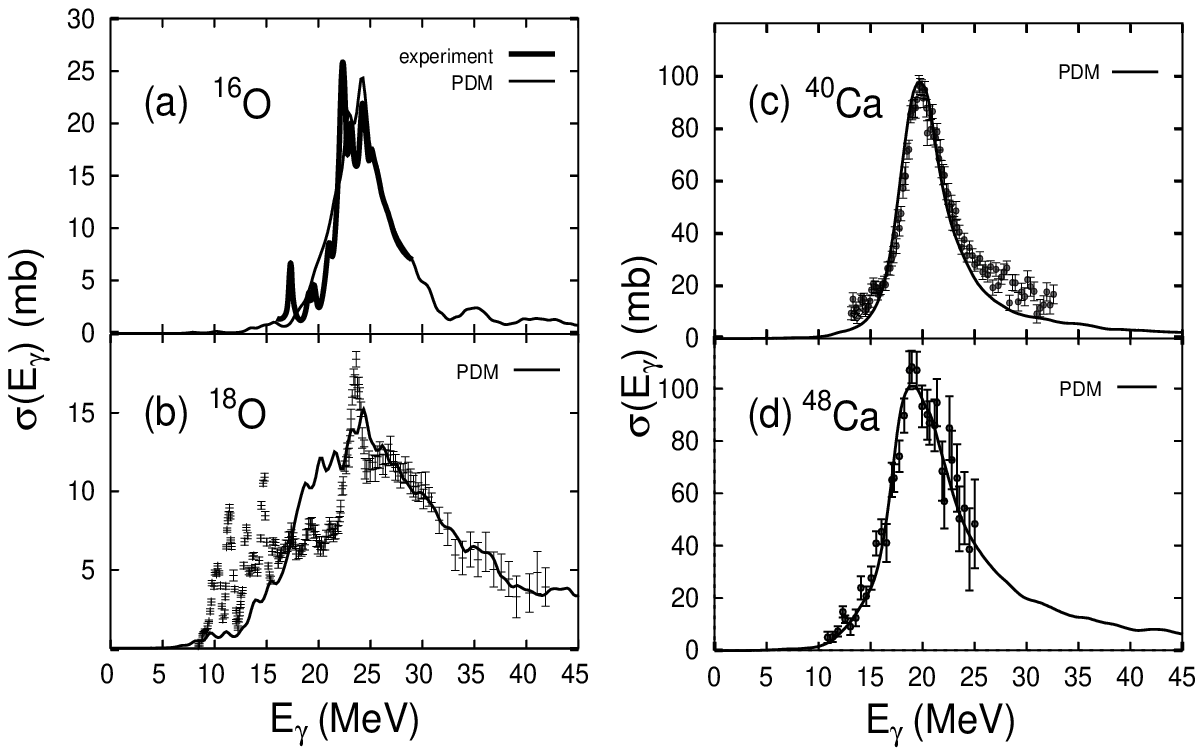}
\caption{Photoabsorption cross sections for $^{16,18}$O and 
$^{40,48}$Ca obtained 
within PDM in comparison with
experimental data~\cite{O18}.\label{PDR}}
\end{figure}
Shown in Fig. \ref{PDR}
are the photoabsorption cross sections $\sigma(E_{\gamma})$, 
which have been obtained within PDM
for $^{16,18}$O and $^{40,48}$Ca~\cite{PDR}. 
The shapes of the calculated photoabsorption 
cross sections are found in overall 
reasonable agreement
with available experimental data~\cite{O18}
The fractions of the energy-weighted sum (EWS) of strength exhausted by the low-energy tail
of GDR are shown in Figs. \ref{EWSR_O}. The trend 
obtained within PDM for oxygen isotopes reproduces the one observed in the 
recent experiments at GSI~\cite{GSI}, which shows a clear deviation 
from the prediction by the cluster sum rule (CSR). 
The agreement between the PDM prediction and the experimental data
for the photoabsorption cross sections as well as for the EWS of 
PDR strength suggests 
that the mechanism of the damping of
PDR is dictated by the coupling between the GDR phonon and 
noncollective $ph$ excitations rather than by the oscillation
of a collective neutron excess against the core. 
Strong pairing correlations also prevent
the weakly bound neutrons to be decoupled from the rest of the 
system~\cite{Mizutori}. 
Only when the GDR is very 
collective so that it  can be well 
separated from the 
neutron excess, the picture of PDR damping becomes closer to the prediction 
by the CM.  
\vspace{2mm}

\hspace{-4mm}
{\bf CONCLUSION}
\vspace{2mm} 

The PDM is a simple yet microscopic model, which can describe rather 
well various resonances and has resolved several long standing problems 
including the width and shape of the hot GDR, the electromagnetic 
cross section of the DGDR, the spreading (quenching) of the GTR. It also predicts 
the PDR in neutron-rich nuclei and the DGDR in hot nuclei.

\end{document}